\documentstyle[12pt,aaspp4]{article}

\received{2000 February 14}

\begin{document}

\title{LOW FREQUENCY RADIO PULSES FROM GAMMA-RAY BURSTS?}

\author{Vladimir V. Usov}

\affil{Department of Condensed-Matter Physics,
Weizmann Institute of Science, Rehovot
76100, Israel}

\author{Jonathan I. Katz}

\affil{Department of Physics and McDonnell Center for the Space Sciences,
Washington University, St. Louis, Mo. 63130}

\authoremail{fnusov@weizmann.weizmann.ac.il, katz@wuphys.wustl.edu}

\date{}

\begin{abstract}
Gamma-ray bursts, if they are generated in the process of interaction
between relativistic strongly magnetized winds and an ambient
medium, may be accompanied by very short pulses of low-frequency 
radio emission. The bulk of this emission is expected to be at
the frequencies of $\sim 0.1-1$ MHz and cannot be observed. However,
the high-frequency tail of the low-frequency radio emission may reach
a few ten MHz and be detected, especially if the strength of 
the magnetic field of the wind is extremely high.

\end{abstract}

\newpage

\section{Introduction}
The prompt localization of gamma-ray bursts (GRBs) by
BeppoSAX led to the discovery of X-ray/optical/radio afterglows and 
associated host galaxies. Subsequent detections of
absorption and emission features at high redshifts ($0.43\leq z \leq 
3.42$) in optical afterglows of GRB and their host galaxies 
clearly demonstrate that at least some of the GRB sources lie at 
cosmological distances (for reviews, see \cite{Piran99}; \cite{Vietri99}).

A common feature 
of all acceptable models of cosmological $\gamma$-ray bursters is that a
relativistic wind is a source of GRB radiation. The Lorentz factor,
$\Gamma_0$, of such a wind is about $10^2-10^3$ or
even more (e.g., \cite{Fenimore93}; \cite{Harding97}).
A very strong magnetic field may be in the plasma 
outflowing from cosmological $\gamma$-ray bursters (\cite{Usov94}; 
\cite{Blackman96}; \cite{Vietri96}; \cite{Katz97}; \cite{Meszaros97}).
A relativistic, strongly magnetized wind interacts with an ambient medium 
(e.g. an ordinary interstellar gas) and decelerates.
It was pointed out (\cite{Meszaros92}) that such an interaction, assumed to
be shock-like, may be responsible for generation of cosmological GRBs. 

The interaction between a relativistic, strongly magnetized wind and
an ambient plasma was studied numerically by Smolsky \& Usov (1996,
2000) and Usov \& Smolsky (1998). These studies showed that
about 70\% of the wind energy is transferred to the ambient plasma
protons that are reflected 
from the wind front. The other $\sim 30$\% of the wind energy losses 
is distributed between high-energy electrons and low-frequency 
electromagnetic waves that are generated at the wind front because of
nonstationarity of the wind---ambient plasma interaction (see below).
High-energy electrons, accelerated at the wind front 
and injected into the region ahead of the front, generate 
synchro-Compton radiation in the fields of the low-frequency 
waves. This radiation closely resembles synchrotron radiation and can 
reproduce  the non-thermal radiation of GRBs observed in 
the Ginga and BATSE ranges (from a few KeV to a few MeV).

\cite{Ginzburg73} and \cite{Palmer93} suggested that GRB might be sources of
radio emission, and that it might be used to determine their distances and,
through the dispersion measure, the density of the intergalactic plasma.
Here we consider some properties of the low-frequency waves generated at the
wind front.  We argue that coherent emission by the high-frequency tail of
these waves may be detected, in addition to the high-frequency (X-ray and
$\gamma$-ray) emission of GRBs, as a short pulse of low-frequency radio
emission (\cite{Katz94,Katz99}).


\section{Low-frequency electromagnetic waves generated at the wind front}
\par
Our mechanism for production of short pulses of low-frequency
radio emission from relativistic, strongly magnetized wind-generated 
cosmological GRBs applies very generally.  For the sake of concreteness, we
consider wind parameters that are natural in a GRB model that involves a
fast rotating compact object like a millisecond pulsar or dense transient
accretion disc with a surface magnetic field $B_s \sim
10^{15}-10^{16}$ G (\cite{Usov92}; \cite{Blackman96}; \cite{Katz97};
\cite{Kluzniak98}; \cite{Spruit99}).

In this model the rotational
energy of compact objects is the energy source of cosmological GRBs.
The electromagnetic torque transfers this energy on a time scale of 
seconds to the energy of a Poynting flux-dominated wind that flows away 
from the object at relativistic speeds, $\Gamma_0\simeq 10^2-10^3$
(e.g., \cite{Usov94}). The wind structure at a time $t\gg\tau_{_\Omega}$
is similar to a shell with radius $r\simeq ct$ and thickness of
$\sim c\tau_{_\Omega}$, where $\tau_{_\Omega}$ is the characteristic
deceleration time of the compact object's rotation, or the dissipation time
of a transient accretion disc.

The strength of the magnetic field at the front of the wind may be as high as

\begin{equation}
B\simeq B_s \frac {R^3}{r_{\rm lc}^2r}
\simeq  10^{15}\frac Rr\left(
\frac{B_{_S}}{ 10^{16}\,\rm{G}}\right) \left(\frac\Omega
{10^4\,\rm{s}^{-1}}\right)^2\rm{G},
\label{Bobj}
\end{equation}
 
\noindent
where $R\simeq 10^6$ cm is the radius of the compact object,
$\Omega\simeq 10^4$ is the angular velocity at the moment of its formation
and $r_{\rm lc}=c/\Omega=3\times 10^6(\Omega /10^4\,$s$^{-1})$~cm
is the radius of the light cylinder. 

The distance at which deceleration of the wind due to its interaction 
with an ambient gas becomes important is

\begin{equation}
r_{\rm{dec}}\simeq 10^{17}\left({\frac{Q_{\rm{kin}}}{10^{53}\,
\rm{ergs}}}\right)^{1/3}\left({\frac n{1\,\rm{cm}^{-3}}}\right)^{-1/3}\left(
{\frac{\Gamma_0}{10^2}}\right)^{-2/3}\rm{cm},
\label{rdec}
\end{equation}

\noindent 
where $n$ is the density of the ambient gas and $Q_{\rm{kin}}$
is the initial kinetic energy of the outflowing wind. 
Eq. (2) assumes spherical symmetry; for beamed flows $Q_{\rm{kin}}$ is
$4\pi$ times the wind energy per steradian.
At $r\sim r_{\rm dec}$, the main part of $Q_{\rm kin}$ is
lost by the wind in the process of its inelastic interaction with the
ambient medium, and the GRB radiation is generated.

Substituting $r_{\rm{dec}}$ for $r$ into equation (\ref{Bobj}), we
have the following estimate for the magnetic field at the wind
front at $r\sim r_{\rm {dec}}$:

\begin{equation}
B_{\rm{dec}}\simeq 10^4 \epsilon_B^{1/2} \left(\frac{B_s}{10^{16}\,\rm{G}}
\right)\left(\frac\Omega{10^4\,\rm{s}^{-1}}\right)^2\left(
\frac{Q_{\rm{kin}}}{10^{53}\,\rm{ergs}}\right)^{-1/3}\left(\frac
n{1\,\rm{cm}^{-3}}\right)^{1/3}\left({\frac{\Gamma_0}{10^2}}\right)
^{2/3}\rm{G}\,,
\label{Bdec}
\end{equation}
 
\noindent
where we have introduced a parameter $\epsilon_B < 1$ which gives the
fraction of the wind power remaining in the magnetic field at the
deceleration radius.  For plausible parameters of cosmological $\gamma$-ray
bursters, $B_{_S} \simeq 10^{16}$~G, $\Omega\simeq 10^4$~s$^{-1}$,
$Q_{\rm{kin}}\simeq 10^{53}$~ergs, $\Gamma_0\simeq 10^2-10^3$ and $n\sim 1$
cm$^{-3}$, from equation (\ref{Bdec}) we have $B_{\rm{dec}}\simeq (1-5)
\times 10^4 \epsilon_B^{1/2}$~G. 

For consideration of the interaction between a
relativistic magnetized wind and an ambient gas, it is convenient
to switch to the co-moving frame of the outflowing plasma (the wind frame).
While changing the frame, the magnetic and electric fields in the wind
are reduced from $B$ and $E=B[1-(1/\Gamma_0^2)]^{1/2}
\simeq B$ in the frame of
the $\gamma$-ray burster to $B_0\simeq B/\Gamma_0$ and $E_0=0$
in the  wind frame.  This is analogous to the the well-known transformation
of the Coulomb field of a point charge: purely electrostatic in the frame of
the charge, but with $E \approx B$ in a frame in which the charge moves
relativistically.  Using this and equation (\ref{Bdec}), for
typical parameters of cosmological $\gamma$-ray bursters we have 
$B_0\simeq (0.5-1)\times 10^2 \epsilon_B^{1/2}$~G at $r\simeq r_{\rm dec}$.

In the wind frame, the ambient gas moves to the wind front with the Lorentz 
factor $\Gamma_0$ and interacts with it. The main parameter which describes 
the wind---ambient gas interaction is the ratio of the energy densities of 
the ambient gas and the magnetic field, $B_0$, of the wind 

\begin{equation}
\alpha = 8\pi n_0m_pc^2(\Gamma_0-1)/B^2_0\,, 
\label{alpha}
\end{equation}

\noindent 
where $n_0=n\Gamma_0$ is the density of the ambient gas in the wind frame
and $m_p$ is the proton mass. 
 
At the initial stage of the wind outflow, $r\ll r_{\rm dec}$, $\alpha \ll
1$, but it increases in the process of the wind expansion as $B_0$
decreases.  When $\alpha$ is $\gtrsim 
\alpha_{\rm cr}\simeq 0.4$ (at $r\sim r_{\rm dec})$, 
the interaction between the wind and the ambient gas is strongly
nonstationary, and effective acceleration of electrons and generation of
low-frequency waves at the wind front both begin  
(\cite{Smolsky96}, 2000; \cite{Usov98}). For $0.4\lesssim \alpha
\lesssim 1$, the mean Lorentz factor of outflowing high-energy electrons
$\Gamma_e^{\rm{out}}$ accelerated at the wind front and the mean field of
low-frequency waves $\langle B_w \rangle$ weakly depend on $\alpha$ (see
Table 1) and are approximately given by

\begin{equation}
\langle\Gamma_e^{\rm {out}}\rangle \simeq 0.2
(m_p/m_e)\Gamma_0 \,\,\,\,{\rm and}\,\,\,\,
\langle B_w\rangle =(\langle B_z\rangle^2 + 
\langle E_y\rangle^2)^{1/2}\simeq 0.1 B_0
\label{Gammamean} 
\end{equation}

\noindent 
to within a factor of 2, where $B_z$ and $E_y$ are the magnetic and
electric field components of the waves. The mechanism of generation
of these waves is coherent and consists of the following:
At the wind front there is a surface 
current that separates the wind matter with a very strong magnetic 
field and the ambient gas where the magnetic field strength is
negligible. This current varies in time because of nonstationarity of
the wind - ambient gas interaction and generates low-frequency waves.

\section{Possible detection}

At $\alpha\sim 1$, for the bulk of high-energy electrons in the region
ahead of the wind front the characteristic time of their synchrotron energy
losses is much less than the GRB duration. In this case, the luminosity
per unit area of the wind front in $\gamma$-rays is
$l_\gamma\simeq m_ec^3n\langle \Gamma_e^{\rm {out}}\rangle$ while the same
luminosity in low-frequency waves is $l_w\simeq c\langle B_w\rangle^2/4\pi$.
Using these, the ratio of the luminosity in low-frequency waves 
and the $\gamma$-ray luminosity is 

\begin{equation}
\delta ={l_w\over l_\gamma}={2\over \alpha}{m_p\over m_e}
{\langle B_w\rangle^2\over B_0^2}
{\Gamma_0\over \langle \Gamma_e^{\rm {out}}\rangle}\,.
\label{delta}
\end{equation}

\par
From Table 1, we can see that the GRB light curves in both 
$\gamma$-rays and low-frequency waves have maximum when $\alpha$ is  
about 0.4, and the maximum flux in low-frequency waves is
about two times smaller than the maximum flux in $\gamma$-rays ($\delta
\simeq 0.5$). At $\alpha > 0.4$ the value of $\delta$ decreases with
increasing $\alpha$. Therefore, we expect that the undispersed duration
$\Delta t_r$ of the low-frequency pulse to be somewhat
smaller than the GRB duration, and the energy
fluence in low-frequency waves to be roughly an order of magnitude smaller
than the energy fluence in $\gamma$-rays.
The rise time of the radio pulse is very short because
that the value of $\langle B_w\rangle^2$ increases very fast when $\alpha$
changes from 0.3 to 0.4 (see Table I).

Figure~1 shows a typical spectrum of low-frequency waves generated at
the wind front in the wind frame. 
This spectrum has a maximum at the frequency $\omega_{\rm max}'$ 
which is about three times higher than the proton
gyrofrequency $\omega_{Bp}=eB_0 /m_pc\Gamma_0$
in the wind field $B_0$. Taking into account the Doppler effect,
in the observer's frame the spectral maximum for
low-frequency waves is expected to be at the frequency

\begin{equation}
\nu_{\rm max}\simeq {2\Gamma_0\over {1+z}}{\omega_{\rm max}'\over 2\pi}
\simeq {eB_0 \over (1+z)m_pc} \simeq {1\over {1+z}}
\left({B_0\over 10^2\,{\rm G}}\right)\,\,\,{\rm MHz}\,,
\label{numax}
\end{equation}
 
\noindent
where $z$ is the cosmological redshift. For typical parameters of
cosmological GRBs, $B_0\simeq (0.5-1)\times 10^2 \epsilon_B^{1/2}$ G and
$z\simeq 1$, we have $\nu_{\rm max}\simeq (0.2-0.5)\epsilon_B^{1/2}$ MHz.
Unfortunately, the bulk of the low-frequency waves is at low frequencies,
and cannot be observed.  However, their high-frequency tail may be detected.

At high frequencies, $\nu > \nu_{\rm max}$,
the spectrum of low-frequency waves may be fitted by a power law
(\cite{Smolsky00}):

\begin{equation}
|B(\nu )|^2\propto 
\nu ^{-\beta}\,,
\label{Bnu}
\end{equation}

\noindent
where $\beta\simeq 1.6$.  In the simulations of the wind---ambient
gas interaction (\cite{Smolsky96}, 2000; \cite{Usov98})
both the total numbers of particles of the ambient gas and the sizes of 
spatial grid cells are restricted by computational reasons, so that the 
the spectrum (\ref{Bnu}) is measured reliably
only at $\nu\lesssim 10\nu_{\rm max}$.  The amplitudes of the computed
oscillations with $\nu > 10\nu_{\rm max}$ are so small ($\lesssim (0.2-0.3)
\langle B_w\rangle$) that they cannot be distinguished from computational
noise (\cite{Smolsky96}).  Future calculations with greater computational
resources may alleviate this problem.

The value of  $\nu_{\rm max}$ depends on many parameters of 
both the GRB bursters and the ambient gas around them, 
and its estimate, $\nu_{\rm max}\simeq (0.2-0.5)\epsilon_B^{1/2}$ MHz, 
is uncertain within a factor of 2--3 or so.  In the most extreme case
in which $\nu_{\rm max}$ is as high as a few~MHz, the high-frequency tail of 
low-frequency waves may be continued up to $\sim 30$ MHz where ground-based
radio observations may be performed. In this case, the energy fluence in a
pulse of radio emission at $\nu\sim 30$ MHz may be as high
as a few percent of the GRB energy fluence in $\gamma$-rays. 

A pulse of low-frequency radio emission is strongly affected by intergalactic 
plasma dispersion in the process of its propagation. At the frequency  
$\nu$, the radio pulse retardation time with respect to a GRB is

\begin{equation}
\tau(\nu)={D\over v} -{D\over c}={e^2 \int n_e\,d\ell \over 2\pi m_ec\nu^2}
\simeq 1.34\times 10^{-3}{\int n_e\,d\ell \over \nu^2}\,\,\,{\rm s}\,,
\label{taunu}
\end{equation}

\noindent
where $\int n_e\,d\ell$ is the intergalactic dispersion measure in
electrons/cm$^2$, $v=cn$ is the group velocity of
radio emission, $n=1-(e^2 n_e/2\pi m_e\nu^2)$ is the refractive index
and $\nu$ in Hz. From equation (\ref{taunu}), for the plausible
parameters of $n_e \simeq 10^{-6}$ cm$^{-3}$ and a distance of $10^{28}$ cm,
at $\nu = 30$ MHz we have $\tau(\nu)\simeq 10^4$ s. This is time
enough to steer a radio telescope for the radio pulse detection. In equation
(\ref{taunu}), we neglected the radio pulse retardation time in our Galaxy,
which is typically one or two orders of magnitude less than that in the
intergalactic gas.

The observed duration of the low-frequency pulse at the frequency $\nu$
in the bandwidth $\Delta \nu$ is

\begin{equation}
\Delta t_{obs}(\nu,\,\Delta \nu )= {\rm max}\,[\Delta t_r,\,2
(\Delta \nu /\nu)\tau (\nu)]\,.
\label{Deltatnu}
\end{equation}

\noindent
For plausible parameters, $\nu \sim 30$ MHz, $\Delta\nu\sim 1$ MHz,
$\tau(\nu )\sim 10^4$ s and $\Delta t_r\sim 1-10^2$ s, we have
$\Delta t_{obs}(\nu,\,\Delta \nu )\sim 7\times 10^2$ s; the observed
duration of low-frequency radio pulses is determined 
by intergalactic plasma dispersion, except for extremely long GRBs. 

It is now possible to estimate, given assumed values for the magnetic field,
the amplitude of the signal produced.  We also assume that the plasma field
couples efficiently to the free space radiation field.  For a radio fluence
${\cal E}_R = \delta {\cal E}_{GRB}$ and a radio fluence spectral density
\begin{equation}
{\cal E}_\nu = \cases{0 & {\rm for}\,\,\,\,
$\nu < \nu_{max}$
,\cr {\beta -1 \over \nu_{max}}
{\cal E}_R \left(
{\nu \over \nu_{max}}\right)^{-\beta} & {\rm for}\,\,\,\,
$\nu \ge \nu_{max}$
,\cr}
\label{Rfluence}
\end{equation}
the radio spectral flux density is
\begin{equation}
F_\nu = \cases{{\delta (\beta - 1) \over \Delta t_r \nu_{max}} \left({\nu
\over \nu_{max}}\right)^{-\beta} {\cal E}_{GRB} & {\rm for}\,\,\,\,
${2 \Delta \nu \over \nu}
\tau(\nu) < \tau_r$
,\cr {\delta (\beta - 1) \over 2 \Delta \nu \tau(\nu)}
\left({\nu \over \nu_{max}}\right)^{1 - \beta} 
{\cal E}_{GRB} & {\rm for}\,\,\,\,
${2 \Delta
\nu \over \nu} \tau(\nu) \ge \tau_r$
.\cr}
\label{Rflux} 
\end{equation}
For the latter (dispersion-limited) case with the parameters ${\cal E}_{GRB}
= 10^{-4}$ erg cm$^{-2}$, $\delta = 0.1\epsilon_B$, $\beta = 1.6$,
$\nu_{max} = 0.3$ MHz, $\nu = 30$ MHz, $\Delta \nu = 1$ MHz, $\tau(\nu) =
10^4$ s we find $F_\nu \approx 2 \times 10^6 \epsilon_B^{(\beta+1)/2}$ Jy.

The appropriate value of $\epsilon_B$ is very uncertain.  In some models it
may be $O(1)$, but for GRB with sharp subpulses its value is limited by the
requirement that the magnetic stresses not disrupt the thinness of the
colliding shells (\cite{Katz97}).  For subpulses of width $\zeta$ of the GRB
width this suggests $\epsilon_B < \zeta^2$; typical estimates are $\zeta
\sim 0.03$ and $\epsilon_B < 10^{-3}$, leading to $F_\nu \lesssim 10^2$ Jy.

These large values of $F_\nu$ may be readily detectable, although the
assumed values of $\epsilon_B$ are very uncertain.   There are additional
uncertainties.  We have assumed that the radio pulse spectrum (11) is valid
up to a frequency $\nu\simeq 30$ MHz that may be hundreds of times higher
than $\nu_{\rm max}$.  As discussed above, the spectrum (11) of
low-frequency waves is calculated directly only at $\nu\lesssim 10
\nu_{\rm max}$. At $\nu > 10 \nu_{\rm max}$, the spectrum must be
extrapolated, with unknown confidence, from the calculations.  The radio
spectral flux density at $\nu\simeq 30$ MHz may therefore be less than the
preceding estimates.  However, even in this case the very high sensitivity
of measurements at radio frequencies may permit the detection of coherent
low-frequency radio emission from GRB.

\section{Discussion} 

In this Letter, we have shown that GRBs may be accompanied by very 
powerful short pulses of low-frequency radio emission. For detection of
these radio pulses it may be necessary to perform observations at lower
frequencies than are generally used in radio astronomy, which are limited by
the problem of transmission through and refraction by the ionosphere.  In
particular, observations from space are free of ionospheric refraction and
are shielded by the ionosphere from terrestrial interference.  Although even
harder to predict, detection from the ground at higher frequencies may also
be possible.

Space observations are possible at frequencies down to that at which the
interstellar medium becomes optically thick to free-free absorption.  This
frequency is
\begin{equation}
\nu_{abs} = 1.0 \times 10^6 \left\langle{n_{0.03}^2 \over T_3^{3/2}}
\right\rangle^{1/2} \vert\csc b^{II}\vert^{1/2}\ {\rm Hz},
\label{ISMabs}
\end{equation}
where $n_{0.03} \equiv n_e/(0.03$ cm$^{-3}$) and $T_3 \equiv
T/10^{3\,\circ}$K (\cite{Spitzer62}), $n_e$ and $T$ are the interstellar
electron density and temperature, respectively, and $b^{II}$ is the Galactic
latitude.  The expression $\langle\rangle$ may be $O(1)$, but could be
substantially larger if the electrons are strongly clumped or cold.
On the other hand, although much of the interstellar volume is filled with
very hot ($10^{6\,\circ}$K) and radio-transparent gas, this probably
contains very little of the electron column density.  Intergalactic
absorption poses a somewhat less restrictive condition if the medium is hot
($10^{6\,\circ}$K), as generally assumed.

At these low frequencies, and even at
tens of MHz, interstellar scintillation (\cite{Goodman97}) will be very
large.  Observations of coherent radio emission from GRB would not only
illuminate the physical conditions in their radiating regions, but would
determine (through the dispersion measure) the mean intergalactic plasma
density and (through the scintillation) its spatial structure.

The flux density implied by Eq. (\ref{Rflux}) appears impressively large,
but it applies only to the brief period when the dispersed signal is
sweeping through the bandwidth $\Delta\nu$ of observation, so that it is
unclear if it is, in fact, excluded by the very limited data
(\cite{Cortiglioni81}) available.  Further, the extrapolation of the
radiated spectrum to $\nu \gg \nu_{max}$ is very uncertain.  Finally, we
have also not considered the (difficult to estimate) temporal broadening of
this brief transient signal by intergalactic scintillation, which will both
reduce its amplitude and broaden its time-dependence.

Searches for radio pulses started about 50 years ago, prior to the discovery
of GRBs. During wide beam studies of ionospheric scintillations, simultaneous
bursts at 45 MHz of $10-20$ s duration were reported by Smith (1950)
at sites 160 km apart. These events were detected at night, approximately
once a week. The origin of these bursts was never determined. 
The results of Smith (1950) were not confirmed by subsequent observations
at frequencies of 150 MHz or higher.

Modern observations (\cite{Dessenne96} at 151 MHz; \cite{Balsano98} at 74
MHz; \cite{Benz98} broadband; see \cite{Frail98} for a review) set upper
bounds to the brightnesses of some GRB at comparatively high frequencies.
These bounds are not stringent, and do not exclude extrapolations of the
lower frequency fluxes suggested here.  There are few data at lower
frequencies.

\acknowledgments

We thank D. M. Palmer for discussions.
V.V.U. acknowledges support from MINERVA Foundation, Munich, Germany.

\clearpage

\begin{table}
\caption{Derived parameters of simulations for both high-energy electron
Lorentz factor $\Gamma_e^{\rm{out}}$ and low-frequency electromagnetic wave
amplitudes $B_w$ and their power ratio $\delta$ in the region ahead of the
wind front}
\label{Results}
$$
\begin{array}{ccccc}
\rm{\alpha} &{\langle\Gamma^{\rm {out}}_e\rangle\over\Gamma_0}&
\langle B_w^2\rangle \over B_0^2  
&  \delta\\
\tableline
\rm {0.2}             & 9.5     &  10^{-5}            & 0.017 \\
\rm {0.3}             & 37      &  1.4\times 10^{-4}  & 0.048 \\
\rm {0.4}             & 396     &  0.021              & 0.49  \\
\rm {0.5}             & 339     &  0.02               & 0.42  \\
\rm {0.57}            & 296     &  0.011              & 0.24  \\
\rm {0.67}            & 276     &  0.008              & 0.16  \\
\rm {1}               & 214     &  0.005              & 0.084 \\
\rm {1.33}            & 137     &  1.6\times 10^{-3}  & 0.032 \\
\rm {2}               & 75      &  4.4\times 10^{-4}  & 0.01  \\
\rm {4}               & 47      &  1.2\times 10^{-4}  & 0.0024 \\
\end{array}
$$

\tablecomments{The accuracy of the derived parameters is $\sim 20$\%
at $0.4\lesssim \alpha\lesssim 2$ and decreases out of this range of
$\alpha$.} 

\end{table}

\clearpage

\bigskip

\noindent
FIGURE CAPTION

\noindent
Fig. 1. Power spectrum of low-frequency electromagnetic waves generated
at the front of the wind in the wind frame in a simulation with
$B_0=300$ G, $\Gamma_0=300$, and $\alpha=2/3$. The spectrum is fitted
by a power law (dashed line).

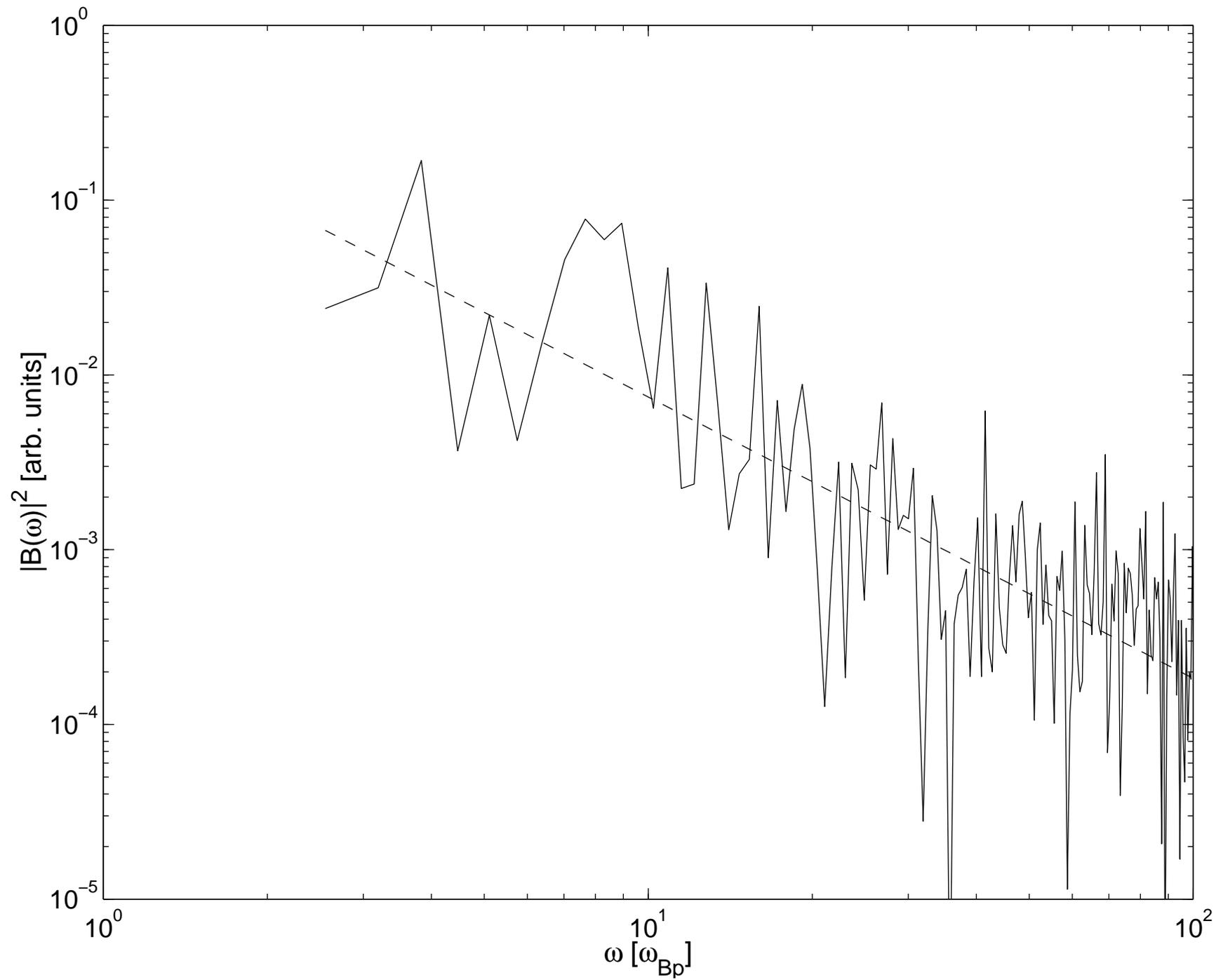
\end{document}